\documentclass[12pt]{article}

\usepackage{epsfig}
\usepackage{graphicx}

\def\be{\begin{equation}}
\def\ee{\end{equation}}
\def\beq{\begin{eqnarray}}
\def\eeq{\end{eqnarray}}

\def\N{{\cal N}}

\def\({\left (}
\def\){\right )}
\def\[{\left [}
\def\[{\right ]}

\def\pr{{(\phi_{,r})}}

\begin{document}

\begin{titlepage}
\bigskip
\rightline{}
\rightline{hep-th/0404261}
\bigskip\bigskip\bigskip\bigskip
\centerline
{\Large \bf {Black Holes with Scalar Hair and }}
\bigskip
\centerline{\Large \bf {Asymptotics in $N=8$ Supergravity}}
\bigskip\bigskip
\bigskip\bigskip

\centerline{\large Thomas 
Hertog\footnote{Hertog@vulcan.physics.ucsb.edu} and 
Kengo Maeda\footnote{kmaeda@kobe-kosen.ac.jp}}
\bigskip\bigskip
\centerline{\em ${}^1$ Department of Physics, UCSB, Santa Barbara, CA 93106}
\bigskip
\centerline{\em ${}^2$ Department of General Education, Kobe City College 
of Technology, }
\centerline{\em 8-3 Gakuen-higashi-machi, Nishi-ku, Kobe 651-2194, Japan}
\bigskip\bigskip
\begin{abstract}

We consider ${\cal N}=8$ gauged supergravity in $D=4$ and $D=5$. 
We show one can weaken the boundary conditions on the metric and on all
scalars with $m^2 < -{(D-1)^2\over 4}+1$ while preserving the asymptotic 
anti-de Sitter (AdS) symmetries. Each scalar admits a one-parameter family of 
AdS-invariant boundary conditions for which the metric
falls off slower than usual. The generators of the asymptotic symmetries 
are finite, but generically acquire a contribution from the scalars. 
For a large class of boundary conditions we numerically find a one-parameter 
family of black holes with scalar hair. These solutions exist above a 
certain critical mass and are disconnected from the Schwarschild-AdS 
black hole, which is a solution for all boundary conditions. 
We show the Schwarschild-AdS black hole has larger entropy than a hairy 
black hole of the same mass. The hairy black holes lift to inhomogeneous 
black brane solutions in ten or eleven dimensions.
We briefly discuss how generalized AdS-invariant boundary conditions 
can be incorporated in the AdS/CFT correspondence.

\end{abstract}

\end{titlepage}

\baselineskip=18pt

\setcounter{equation}{0}
\section{Introduction}

Stationary, asymptotically flat, vacuum black holes in four dimensions are 
completely characterized by their mass and angular momentum \cite{Israel67}
and have horizons of spherical topology \cite{Hawking72}. But there 
appears to be a much
richer spectrum of black hole solutions in higher dimensional spacetimes.
Recently, a solution of the Einstein vacuum equations in five dimensions 
was found that describes a rotating black ring with a horizon of topology
$S^1\times S^2$ \cite{Emparan02}. Since black rings can carry the same 
asymptotic charges as the rotating Myers-Perry black holes \cite{Myers86} this
means the uniqueness theorem for stationary black holes does not extend to 
five dimensions.

In higher dimensional spacetimes with compactified extra dimensions, there 
exist simple vacuum solutions describing black $p$-branes that are
translationally invariant along the compact directions. Gregory and Laflamme 
(GL) showed these uniform black $p$-branes are unstable below a critical mass 
\cite{Gregory93}. More recently, a branch of non-uniform black string 
solutions 
emerging from the critical uniform string was constructed numerically, first 
in perturbation theory \cite{Gubser02} and then non-perturbatively in
\cite{Wiseman03}. In addition, it has been argued there exists yet another 
family of 
static non-uniform black string solutions that are the end state of the decay
of unstable uniform strings \cite{Horowitz01}. These results indicate that the 
black hole uniqueness theorems do not hold in higher-dimensional, 
asymptotically flat spacetimes with compactified extra dimensions. 

The GL instability persists with asymptotically anti-de Sitter (AdS) boundary 
conditions. In \cite{Hubeny02} the threshold unstable mode was identified 
for small Schwarzschild-$AdS_5\times {\bf S^5}$ black holes, which are 
solutions to type IIB supergravity. This instability
was interpreted as signaling a phase 
transition in the dual gauge theory. Four dimensional AdS-Reissner-Nordstrom 
black holes, which are solutions of ${\cal N}=8$ gauged supergravity, exhibit 
a somewhat similar instability precisely 
when they are locally thermodynamically 
unstable \cite{Gubser01}. This suggests that the black hole uniqueness theorems
do not hold in asymptotically anti-de Sitter spacetimes with compactified
extra dimensions. One of the objectives of this paper is to demonstrate this
explicitely. We will do so by numerically constructing a new class of static 
black brane solutions in 11-dimensional supergravity compactified on $S^7$ 
and in 10-dimensional type IIB supergravity on $S^5$.

M-theory on $S^7$ can be consistently truncated to ${\cal N}=8$ $D=4$ gauged 
supergravity in four dimensions \cite{deWit87}. Similarly $\N=8$ $D=5$ 
supergravity is believed to be a consistent truncation of ten dimensional 
type IIB supergravity on $S^5$. Because of the scalar potential introduced by 
the gauging procedure the maximally supersymmetric vacuum solutions are 
$AdS_4$ and $AdS_5$. Therefore an appealing way to try to find ten or eleven 
dimensional black brane solutions is to look for asymptotically AdS black holes
with scalar hair in the dimensionally reduced supergravity theories.

The original no hair theorem of Bekenstein \cite{Bekenstein74} proves there
are no asymptotically flat black hole solutions with scalar hair for minimally 
coupled scalar fields with convex potentials. This result was extended to the 
case of minimally coupled scalar fields with arbitrary positive potentials in 
\cite{Heusler92}. Later it was shown 
\cite{Sudarsky02} there are no hairy, asymptotically AdS black holes where 
the scalar field asymptotically tends to the true minimum of the potential. In 
\cite{Torii:2001pg}, however, an example was given of a hairy black hole where 
the scalar field asymptotically goes to a negative maximum of the potential.
But because the mass of this solution diverges it is not obvious one can 
regard it as being asymptotically AdS in a meaningful way. 

More recently, a one-parameter family of hairy AdS black holes was
found in three dimensions \cite{Henneaux02}. Asymptotically the scalar field 
again tends to a negative maximum but the potential satisfies the 
Breitenlohner-Freedman (BF) bound \cite{Breitenlohner82}. 
A careful analysis of the asymptotic solutions revealed 
they preserve the asymptotic AdS symmetry group
\cite{Henneaux02}, despite the fact that the standard gravitational mass 
diverges. The reason is that the generators of the asymptotic symmetries
acquire a contribution from the scalar field, which renders the
conserved charges finite. Therefore at 
least in three dimensional gravity coupled to a single scalar field 
(with $m^2=-3/4$), the usual set of AdS-invariant boundary conditions - which 
corresponds to requiring finite gravitational mass - does not include all 
asymptotically AdS solutions. The results of \cite{Henneaux02} 
indicate there are theories that admit a much larger class of AdS-invariant 
boundary conditions than those which have been considered so far.
This also raises the possibility there is a scalar no hair theorem for some
asymptotically AdS boundary conditions and not for others.
We investigate these issues in this paper.

Generalized AdS-invariant boundary conditions are studied in section 2. 
We show that the results of \cite{Henneaux02} generalize to $d$-dimensional
gravity minimally coupled to a scalar field with arbitrary mass $m^2$ in the 
range 
$-{(d-1)^2\over 4} \leq m^2 < -{(d-1)^2\over 4}+1$. In particular we show
there is a one-parameter family of boundary conditions on the scalar and the 
metric components 
that preserve AdS invariance with well defined generators of the 
asymptotic symmetries. For all AdS-invariant boundary conditions except one or
two, the metric as well as the scalar field fall off slower than usual. 
In section 3 we turn to ${\cal N}=8$ $D=5$ gauged supergravity, which 
contains scalars saturating the BF bound. We first write down the generalized 
AdS-invariant boundary conditions on the metric components and the scalars.
For a large class of AdS boundary conditions we then numerically find a 
one-parameter family of black hole solutions with scalar hair.
When lifted to ten dimensions, these solutions describe electro-vacuum black 
branes with a perturbed five sphere on the horizon. In section 4 we consider 
${\cal N}=8$ $D=4$ supergravity and find hairy black holes for generalized 
AdS boundary conditions on scalars above the BF bound.
In section 5, which is reasonably self-contained, we discuss how generalized 
AdS-invariant boundary conditions can be incorporated in the AdS/CFT 
correspondence \cite{Maldacena97}. Finally, in section 6 we summarize our 
results.
\newline
$Note:$Today, the paper \cite{Henneaux04} appeared on hep-th, which 
contains results that overlap with some of those presented in section 2 of 
this paper.

\section{Asymptotically AdS spaces with non-localized matter}

\subsection{Tachyonic Scalars in AdS}

Recall that if we write $AdS_d$ in global coordinates
\be \label{adsmetric}
ds^2_0 = \bar g_{\mu \nu} dx^{\mu} dx^{\nu}=
-(1+{r^2 \over l^2})dt^2 + {dr^2\over 1+r^2/l^2} + r^2 d\Omega_{d-2}
\ee
then for $m^2 <0$ solutions to $\nabla^2\phi -m^2\phi=0$ with harmonic time
dependence $e^{-i\omega t}$ all fall off asymptotically like 
\be\label{genfall}
\phi = {\alpha \over r^{\lambda_{-}}}  + {\beta \over r^{\lambda_{+}}}
\ee
where
\be\label{fallofftest}
\lambda_\pm = {d-1 \pm \sqrt{(d-1)^2 + 4l^2 m^2}\over 2}
\ee
The BF bound is
\be \label{BF}
m_{BF}^2 = -{(d-1)^2\over 4l^2}.
\ee
For fields which saturate this bound, $\lambda_+ = \lambda_- \equiv \lambda$ 
and the second solution asymptotically behaves like $\ln r/r^{\lambda}$.

To have a definite theory one must impose boundary conditions on the 
timelike boundary at spacelike infinity. For reflective boundary conditions 
$\alpha=0$ it is well known that a scalar field with negative mass 
squared does not cause an instability in anti de Sitter space, provided that 
$m^2 \ge  m^2_{BF}$ \cite{Breitenlohner82,Mezincescu85}. This is important, 
since many supergravity theories arising in the low energy limit of string 
theory contain fields with negative $m^2$, but they all satisfy this bound.
For those boundary conditions \cite{Hawking83}
there is a positive energy theorem \cite{Abbott82,Gibbons83,Townsend84} which 
ensures that the total energy cannot be negative whenever this condition is 
satisfied. 

But for $m^2_{BF} \leq m^2 <m^2_{BF} +1$ 
both solutions (\ref{fallofftest}) are 
normalizable. It has been argued 
\cite{Breitenlohner82,Balasubramanian98,Klebanov99} that
scalars with masses in this
range allow a second AdS-invariant quantization that corresponds to choosing 
$\beta=0$ in (\ref{genfall}). However, it is easy to see
this choice cannot define a quantum field theory on the usual AdS background 
(\ref{adsmetric}). 

The standard set of boundary conditions on the metric 
components at spacelike infinity that is left invariant under $SO(d-1,2)$
is given by \cite{Henneaux85}
\beq \label{usual}
g_{rr} = {l^2 \over r^2} -{l^4 \over r^4} +O(1/r^{d+1})  
& \qquad \ \ \ \ \ \ \ \ \ \ \ \ 
g_{tt}= -{r^2 \over l^2} -1 +O(1/r^{d-3})\nonumber\\
g_{tr}=O(1/r^{d})\ \ \  & \qquad g_{ra}=O(1/r^{d})\ \  \ \ \ \nonumber\\
g_{at}=O(1/r^{d-3}) & \qquad \ \ \ \ \ g_{ab}=\bar g_{ab}+O(1/r^{d-3})
\eeq
where $a,b$ label the angular coordinates on $S^{d-2}$.
A generic asymptotic Killing vector field $\xi^\mu$ behaves as
\beq 
\xi^r&=&O(r)+O(r^{-1}), \nonumber \\
\xi^t&=&O(1)+O(r^{-2}), \nonumber\\
\xi^a&=&O(1)+O(r^{-2})
\eeq	
and the charges that generate the asymptotic symmetries are given by
\be\label{gravch}
Q_{G}[\xi]=\frac{1}{2}\oint dS_i
\bar G^{ijkl}(\xi^\perp \bar{D}_j h_{kl}-h_{kl}\bar{D}_j\xi^\perp)
+2\oint dS_i\frac{\xi^i {\pi^i}_j}{\sqrt{\bar g}}
\ee
where $G^{ijkl}={1 \over 2} g^{1/2} (g^{ik}g^{jl}+g^{il}g^{jk}-2g^{ij}g^{kl})$,
$h_{ij}=g_{ij}-\bar{g}_{ij}$ is the deviation from the spatial metric 
$\bar{g}_{ij}$ of pure AdS and $\bar{D}_i$ denotes covariant differentiation 
with respect to $\bar{g}_{ij}$. 

Consider now a simple time symmetric, spherical test field
configuration $\phi$ on an initial time slice, so that $\phi <<1$ everywhere.
The coefficient $M$ of the $O(1/r^{d+1})$ correction to the asymptotic behavior
of the $g_{rr}$ component is then proportional to the total mass
$Q_{G}[\partial_{t}]$ of the scalar field configuration. The constraint equation 
yields
\be
M= Q_{G}[\partial_{t}]= 
{(d-2)\pi^{{d-1 \over 2}} \over 2\Gamma \left({d-1 \over 2} \right)} 
\lim_{r \rightarrow \infty} \int_0^{r} \left(m^2 \phi^2 +(D \phi)^2 \right)
\tilde r^{d-2} d\tilde r.
\ee
However one finds 
this diverges for $\beta=0$ boundary conditions for all $m^2$ in the  
range $m^2_{BF} \leq m^2 <m^2_{BF} +1$, even for arbitrarily small fields. 
It is, therefore, inconsistent to quantize a test field with these falloff
conditions on the standard anti-de Sitter background (\ref{adsmetric}).
More generally, solutions with $\beta=0$ boundary conditions on the scalar
field cannot be asymptotically anti-de Sitter in the usual sense (\ref{usual}).
Of course, this does not exclude the possibility that $\beta=0$ is a valid
scalar field boundary condition on a different AdS background, in which the 
asymptotic behavior of the metric 
is somehow relaxed whilst preserving the asymptotic AdS symmetry group with
well defined generators. This is the subject of the next subsection.

\subsection{Generalized AdS-invariant boundary conditions}

We define asymptotically anti-de Sitter spacetimes by a set of boundary
conditions at spacelike infinity which satisfy the requirements set out in 
\cite{Henneaux85}. We first consider $d \geq 3$ dimensional gravity minimally 
coupled to a self-interacting massive scalar field with 
$ m^2_{BF}<m^2 <1+m^2_{BF}$. We return below to the case in which the BF
bound is saturated. The Hamiltonian is given by
\beq \label{ham}
H[\xi]&=&\int d^{d-1}x \xi^\mu{\cal H}_\mu(x) 
+Q_\phi[\xi]+Q_G[\xi] \nonumber \\
& = &\int d^{d-1}x(\xi^{\perp}{\cal H}_\perp(x)+\xi^i {\cal H}_i(x))+
Q_\phi[\xi]+Q_G[\xi]
\eeq
The ${\cal H}_\mu$ are the usual Hamiltonian and momentum constraints, 
\beq 
{\cal H}_\perp & = &\frac{2}{\sqrt{g}}(\pi^{ij}\pi_{ij}-
\frac{\pi^2}{d-2}+\frac{p^2}{4})
+\sqrt{g}\left[-\frac{R}{2}+\frac{1}{2}(D\phi)^2+V(\phi)\right],
\nonumber \\
{\cal H}_i & = & -2\sqrt{g}D_j\left(\frac{{\pi^j}_i}{\sqrt{g}}\right)
+pD_i\phi.
\eeq
where $\pi^{ij}$ and $p$ are the momenta conjugate to 
$g_{ij}$ and $\phi$. Here we have set $8\pi G=1$.  
The requirement that the Hamiltonian (\ref{ham}) should have well defined
functional derivatives determines the variation of the surface integrals
\cite{Regge74},
\be
\label{scalar-energy}
\delta Q_\phi[\xi]=-\oint dS_i\delta \phi
\left[D^i\phi\,\xi^\perp +\frac{p\xi^i}{\sqrt{g}}\right]   
\ee 
and
\be 
\label{gravitational energy}
\delta Q_G[\xi]=\frac{1}{2}\oint dS_i
G^{ijkl}(\xi^\perp D_j\delta g_{kl}-\delta g_{kl}D_j\xi^\perp)
+2\oint dS_i \frac{\xi^j\delta {\pi^i}_j}{\sqrt{g}}
-\oint dS_i\xi^i\frac{\pi^{jk}\delta g_{jk}}{\sqrt{g}}. 
\ee

To have a definite theory one must impose $some$ boundary conditions
at spacelike infinity. This means $\beta$ should generally be some function
of $\alpha$ in (\ref{genfall}). Consider now the class of solutions 
with the following asymptotic behavior,
\be\label{falloff4dphi}
\phi (r,t,x^a) = {\alpha (t,x^a) \over r^{\lambda_{-}}} +
{f \alpha^{\lambda_{+}/\lambda_{-}} (t,x^a) \over r^{\lambda_{+}}}
\ee
\beq \label{falloff4dg}
g_{rr}=\frac{l^2}{r^2}-\frac{l^4}{r^4}
-{\alpha^2 l^2\lambda_{-} \over (d-2)r^{2+2\lambda_{-}}}
+O(1/r^{d+1})\ \  
& \quad g_{tt}=-{ r^2 \over l^2} -1+O(1/r^{d-3}) \nonumber\\
g_{tr}=O(1/r^{d-2}) \qquad \qquad \qquad & \ 
g_{ab}= \bar g_{ab} +O(1/r^{d-3}) 
\nonumber\\
g_{ra} = O(1/r^{d-2}) \qquad \qquad \qquad & g_{ta}=O(1/r^{d-3}) \ \ \ \ \ \  
\eeq
where $f$ is an arbitrary constant without variation.
When $f=0$ we recover the $\beta=0$ boundary conditions discussed above. 
The standard $\alpha=0$ boundary conditions for localized matter distributions
are obtained for $f \rightarrow \infty$ 
(together with $\alpha \rightarrow 0$). 
Remarkably, for all values of $f$ this set of boundary conditions 
preserves the asymptotic anti-de Sitter symmetries. Thus there exists a 
one-parameter family of AdS-invariant boundary conditions, parameterized 
by $f$.

For $f \rightarrow \infty$ the asymptotic conditions (\ref{falloff4dg}) 
on the metric components reduce to the standard set (\ref{usual}).
The variation of the gravitational charges $\delta Q_G[\xi]$ is finite
in this case, yielding finite conserved charges given by (\ref{gravch}).
The scalar charges are zero, as one expects from localized matter. 

On the other hand, for all finite $f$ both $\delta Q_G[\xi]$ and
$\delta Q_{\phi}[\xi]$ diverge like $r^{d-1-2\lambda_-}$.
The divergences, however, precisely cancel out. The total charge
can therefore be integrated\footnote{The boundary conditions on $\pi^{ij}$ are 
$\pi^{rr}=O(1/r)$, $\pi^{ra}=O(1/r^2)$ and 
$\pi^{ab}=O(1/r^{d-6-2\lambda_{-}})$. Hence the third term in eq.~(\ref{gravitational energy})
is zero. The second term in eq.~(\ref{scalar-energy}) also vanishes because
$p\sim r^{d-3-2\lambda_-}$.}, giving
\beq
\label{charge4d}
Q[\xi]&=&Q_G[\xi]+{\lambda_{-} \over 2}\oint d\Omega_{d-2}\frac{\xi^\perp}{r}
r^{d-1}\left( \phi^2 +{2f(\lambda_{+} - \lambda_{-}) \over d-1}
\phi^{\frac{d-1}{\lambda_-}} \right)\nonumber\\
& = & \tilde Q_G[\xi]+\frac{2f\lambda_-\lambda_+}{d-1}
\oint d\Omega_{d-2}\frac{\xi^\perp}{r}
r^{d-1}\phi^{\frac{d-1}{\lambda_-}} 
\eeq
where $d\Omega_{d-2}$ is the volume element on the unit $d-2$ sphere
and $\tilde Q_G[\xi]$ is the finite part of the gravitational 
charge, coming from the standard asymptotic corrections to the AdS metric.

We emphasize again that in the theory defined by $f=0$ boundary conditions, 
which is 
often used in AdS/CFT, one must relax the asymptotic falloff of some metric 
components to ensure backreaction can be made small and the asymptotic 
AdS symmetry group is preserved. Although there is no residual finite
scalar contribution to the total charges $Q$ in this case, it is only the
variation of the sum of both charges that is well defined.

Finally we turn to the case in which the BF bound is saturated.
The second independent solution of the linearized scalar field equation
now asymptotically behaves like $\ln r/r^{\lambda}$. The logarithmic component
somewhat alters the formulas but there is no essential difference - there is
again a one-parameter family of AdS-invariant boundary conditions.
Indeed, we find the asymptotic AdS symmetry group is preserved for
solutions with the following asymptotic behavior,
\be 
\phi = {\alpha \over r^{\lambda}} \ln r +
{\alpha \over r^{\lambda}}\left(f-{1 \over \lambda} \ln \alpha \right) 
\ee
\be\label{falloff5dg}
g_{rr} = \frac{l^2}{r^2} -\frac{l^4}{r^4}
-{\alpha^2 l^2 \lambda (\ln r)^2 \over (d-2)r^{d+1}}
-{\alpha^2 l^2(2\lambda f-1-2\ln \alpha)  \over (d-2) } 
{\ln r \over r^{d+1}} +O(1/r^{d+1})
\ee
\beq
g_{tr}=O(1/r^{d-2}), 
& \qquad g_{ab} = \bar g_{ab} +O(1/r^{d-3}),
& \qquad g_{ta}=O(1/r^{d-3}) \nonumber\\
g_{ar}=O(1/r^{d-2}),
& \qquad \ \ \ \ \ \ g_{tt}=-\frac{r^2}{l^2}-1+O(1/r^{d-3}) 
\eeq
where $\alpha(t,x^a)$ and $f$ is again an arbitrary constant.
For finite $f$, the variations of the gravitational and scalar charges 
are logarithmically divergent. But the divergences again cancel out,
allowing us to integrate the total charges $Q=Q_\phi +Q_G$. 
This yields\footnote{The boundary conditions on the momenta are
$\pi^{rr}=O(1/r)$, $\pi^{ra}=O(1/r^2)$ and
$\pi^{ab}=O(\ln^2 r/r^5)$.}
\be
\label{charge5d}
Q[\xi] = \tilde Q_G[\xi]+{1 \over 2} \oint d\Omega_{d-2} \frac{\xi^\perp}{r} 
\left(\lambda \beta^2 -\alpha \beta +\frac{\alpha^2}{2\lambda}\right)
\ee
where $\beta = \alpha(f- \ln \alpha/\lambda)$.
\newline
We will use this expression and (\ref{charge4d}) in the next sections
to compute the mass of the hairy black hole solutions we will find.

For $f \rightarrow \infty$ we recover the usual 
falloff conditions on the metric components. Even though the 
logarithmic mode is switched off in this case, there is still a finite scalar 
contribution to the conserved charges.  This is also evident in 
the spinorial proof \cite{Gibbons83} of the positive energy theorem, 
where the positive Nester mass (which equals $Q[\xi]$) contains an extra 
scalar contribution \cite{Hertog03b}.
It means the standard gravitational mass that appears in the metric can 
be negative and need not be conserved during evolution.
In other words, for scalar fields saturating the BF bound positivity of the 
gravitational mass requires boundary conditions that are 
stronger than those required for finite mass.

\setcounter{equation}{0}
\section{Hairy Black Holes in ${\cal N}=8$ $D=5$ Supergravity}

\subsection{AdS-invariant boundary conditions}

${\cal N}=8$ gauged supergravity in five dimensions
\cite{Gunaydin85,Pernici85} is thought to be a consistent 
truncation of ten dimensional type IIB supergravity on $S^5$. The spectrum of 
this compactification involves 42 scalars parameterizing the coset 
$E_{6(6)}/USp(8)$. The scalars that are important for our discussion saturate 
the BF bound and correspond to the subset that parameterizes the coset 
$SL(6,R)/SO(6)$. From the higher dimensional viewpoint, these arise from the 
$\ell =2 $ modes on $S^5$. The relevant part of the action 
involves five scalars $\phi_i$ and takes the form 
\be\label{act}
S =  \int \sqrt{-g} \left [\frac{1}{2}  R -
\sum_{i=1}^{5}\frac{1}{2}(\nabla \phi_{i})^2 -V(\phi_{i})\right ]
\ee
where we have set $8\pi G=1$. The potential for the scalars $\phi_{i}$ is 
given in terms of a superpotential $W(\phi_i)$ via
\be \label{superpot}
V = \frac{g^2}{4} \sum_{i=1}^5 
    \left( \frac{\partial W}{\partial\phi_i} \right)^2 - 
    \frac{g^2}{3} W^2 \qquad ,
\ee
$W$ is most simply expressed as
\be
W =- \frac{1}{2\sqrt{2}} \sum_{i=1}^6 e^{2\beta_i}
\ee
where the $\beta_i$ sum to zero, and are related to the five $\phi_{i}$'s
with standard kinetic terms as follows,
\be\label{alphamatrix}
\pmatrix{ \beta_1 \cr \beta_2 \cr \beta_3 \cr \beta_4 \cr \beta_5 \cr 
\beta_6 } = 
\pmatrix{ 1/2 & 1/2 & 1/2 & 0 & 1/2\sqrt{3} \cr
              1/2 & -1/2 & -1/2 & 0 & 1/2\sqrt{3} \cr
              -1/2 & -1/2 & 1/2 & 0 & 1/2\sqrt{3} \cr
              -1/2 & 1/2 & -1/2 & 0 & 1/2\sqrt{3} \cr
              0 & 0 & 0 & 1/\sqrt{2} & -1/\sqrt{3} \cr
              0 & 0 & 0 & -1/\sqrt{2} & -1/\sqrt{3} }
\pmatrix{ \phi_1 \cr \phi_2 \cr \phi_3 \cr \phi_4 \cr \phi_5 }
\ee
The potential reaches a negative local maximum when all the scalar
fields $\phi_{i}$ vanish. This is the maximally supersymmetric AdS state,
corresponding to the unperturbed $S^5$ in the type IIB theory.
At linear order around the AdS solution, the five scalars each obey the
free wave equation with mass
\be
m^2_{i}= -4
\ee
which saturates the BF bound (\ref{BF}) in five dimensions. Therefore
it is trivial to generalize the results of the previous section
to include more than one scalar.
One finds asymptotic AdS invariance is preserved for solutions with
the following asymptotic behavior, 
\be \label{phi5d}
\phi_{i}(r,t,x^a)=\frac{\alpha_{i}(t,x^a)}{r^2}\ln r +
\frac{\alpha_{i}(t,x^a)}{r^2}\left( f_{i}-{1 \over 2} \ln \alpha_{i}\right)
\ee
\be \label{aads5d}
g_{rr}={1 \over r^2} -{1 \over r^4} -\sum_{i=1}^{5}
\frac{2\alpha^2_{i}}{3r^6}(\ln r)^2
-\sum_{i=1}^{5}
\frac{2\alpha^2_{i}}{3r^6}\left(2f_{i}-\ln \alpha_{i}-1/2\right)\ln r
+O(1/r^6)\nonumber\\ 
\ee
\beq
g_{tr}=O(1/r^{3}), 
& \qquad g_{ab} = \bar g_{ab} +O(1/r^{2}),
& \qquad g_{ta}=O(1/r^2) \nonumber\\
g_{ar}=O(1/r^{3}),
& \qquad \ \ \ \ \ \ g_{tt}=-\frac{r^2}{l^2}-1+O(1/r^2) 
\eeq
where $x^a=\chi,\theta,\phi$ and $f_i$ are five constants labelling the
different boundary conditions. The charges $Q$ are given by 
\be
\label{charge5dN8}
Q[\xi]=Q_G[\xi]+{1 \over 2} \sum_{i=1}^5
\oint d\Omega_{d-2} \frac{\xi^\perp}{r} r^{d-1}
\left(\lambda \phi^2_{i} -{\alpha_{i} \phi_{i} \over r^{\lambda}} +
\frac{\alpha^2_{i}}{2\lambda r^{2\lambda}}\right),
\ee
which is finite for a generic asymptotic Killing vector field.

Nonperturbatively, the five scalars $\alpha_{i}$ couple to each other and 
it is generally not consistent to set only some of them to zero.
The exception is $\alpha_{5}$, which does not act as a source for any of 
the other fields. In the next 
section we will consider solutions involving only 
$\alpha_5$, so $\alpha_{i}=0$, $i=1,..4$. 
Writing $\alpha_5 = \phi$ and setting $g^2=4$ so that the AdS radius 
is equal to one, the action (\ref{act}) further reduces to 
\be \label{lagr}
S = \int \sqrt{-g}\left[ \frac{1}{2} R -\frac{1}{2}(\nabla \phi )^2 
+\(2e^{2\phi/\sqrt{3}}_{\ }+4e^{-\phi/\sqrt{3}}_{\ } \)\right ]
\ee

\subsection{Black Holes with Scalar Hair}

We now numerically solve the field equations derived from
(\ref{lagr}) to find a class of static, 
spherically symmetric black hole solutions with scalar hair outside the 
horizon. Writing the metric as
\be 
ds_5^2=-h(r)e^{-2\delta(r)}dt^2+h^{-1}(r)dr^2+r^2d\Omega_3^2. 
\ee
The Einstein equations read
\be \label{field5d1}
h\phi_{,rr}+\left(\frac{3h}{r}+\frac{r}{3}\phi_{,r}^2h+h_{,r} 
\right)\phi_{,r} =V_{,\phi}, 
\ee
\be \label{field5d2}
2(1-h)-rh_{,r}-\frac{r^2}{3}\phi_{,r}^2h=\frac{2}{3}r^2V(\phi),  
\ee
\be \label{field5d3}
\delta_{,r}=-\frac{r}{3}\phi_{,r}^2
\ee
Regularity at the event horizon $R_e$ imposes the constraint
\be \label{horcon}
\phi'(R_{e}) = {R_{e}V_{,\phi_{e}} \over 2-2R_{e}^2V(\phi_{e})/3}
\ee
\begin{figure}[htb]
\begin{picture}(0,0)
\put(47,254){$\phi_e$}
\put(411,19){$R_{e}$}
\end{picture}
\mbox{\epsfxsize=14cm \epsfysize=9cm \epsffile{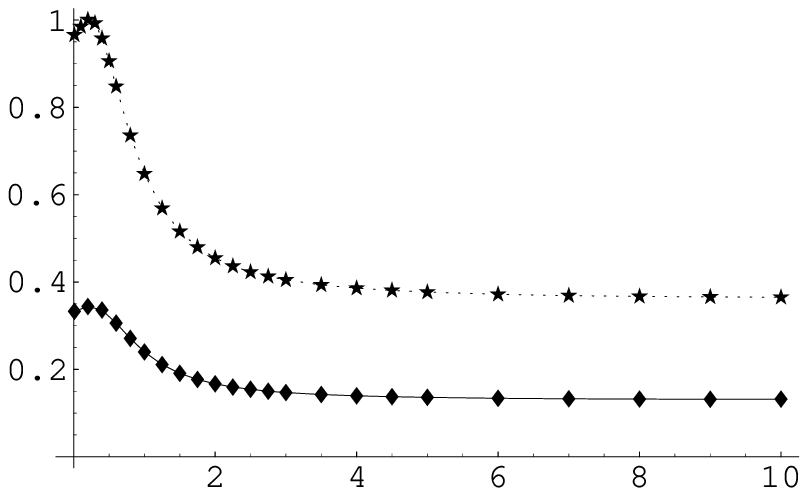}}
\caption{The scalar field $\phi_e$ at the horizon as a function of horizon size
$R_e$ in hairy black hole solutions of $D=5$ ${\cal N}=8$ supergravity.
The two curves correspond to solutions with two different AdS-invariant 
boundary conditions, namely $f=0$ (bottom) and $f=1$ (top).}
\label{1}
\end{figure}
Asymptotic AdS invariance requires $\phi$ asymptotically decays as
\be \label{hair5d}
\phi(r)=\frac{\alpha}{r^2}\left( \ln r-\frac{1}{2}\ln \alpha 
+f \right), 
\ee
where $f$ is a constant whose value is determined by the boundary
conditions. Hence asymptotically
\be \label{asmetric5d}
h(r)=r^2+1+\frac{2\alpha^2}{3r^2}(\ln r)^2+\frac{2\alpha^2}{3r^2}
\left(2f-{1 \over 2}-\ln \alpha \right)\ln r-\frac{M_0}{r^2}, 
\ee
where $M_0$ is an integration constant.

\begin{figure}[htb]
\begin{picture}(0,0)
\put(34,125){$\phi$}
\put(204,14){$r$}
\put(250,141){$r^2\phi/\ln r$}
\put(455,14){$r$}
\end{picture}
\mbox{
\epsfxsize=7cm
\epsffile{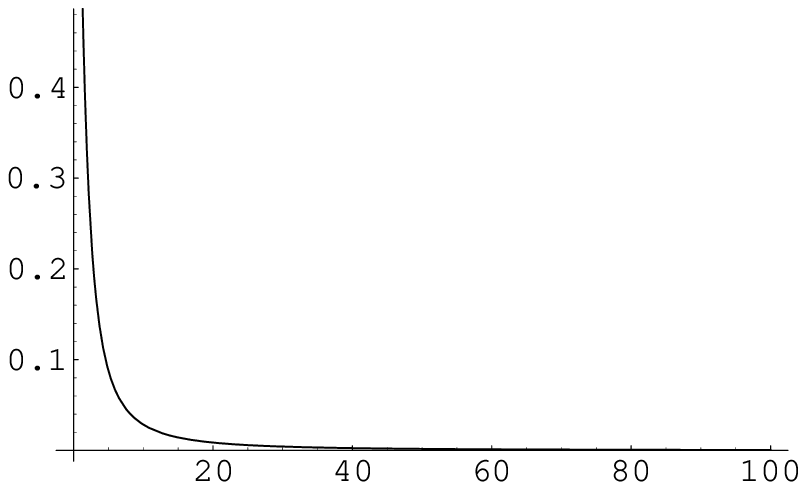}
\raisebox{2.3cm}{~~~~
\begin{minipage}{10cm}
\epsffile{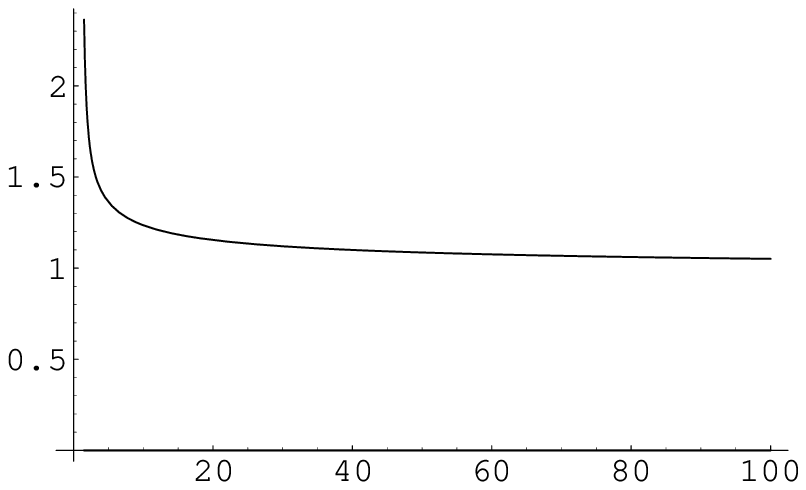}
\end{minipage}}}
\caption{The hair $\phi(r)$ (left) and $r^2\phi/\ln r$ (right) outside a 
black hole of size $R_e=.2$, with boundary conditions specified by $f=1$.}
\label{2}
\end{figure}

The Schwarschild-AdS black hole with $\phi=0$ everywhere
is clearly a solution for all
boundary conditions. The conserved charge (\ref{charge5dN8}) reduces to
\be\label{mass5dschw}
Q[\partial_{t}]=Q_G[\partial_{t}]=3\pi^2 M_0 = 3\pi^2 (R_e^4 +R_e^2),
\ee
which is the usual Schwarschild-AdS mass.
However, numerical integration of the field equations 
(\ref{field5d1})-(\ref{field5d2}) shows that all boundary conditions 
corresponding to finite $f$ also admit a one-parameter family of 
static spherically symmetric black hole solutions with scalar hair outside 
the horizon.

\begin{figure}[htb]
\begin{picture}(0,0)
\put(50,250){$\alpha$}
\put(411,19){$R_{e}$}
\end{picture}
\mbox{\epsfxsize=14cm \epsfysize=9cm \epsffile{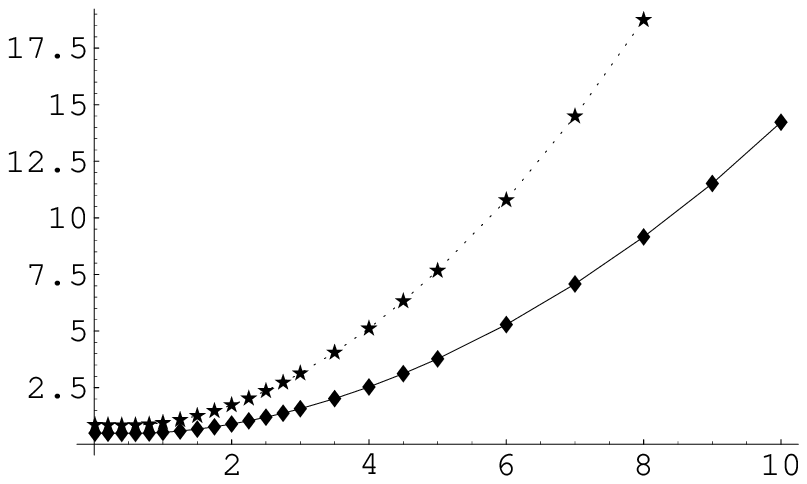}}
\caption{The coefficient $\alpha$ that characterizes the asymptotic profile 
of the hair $\phi(r)$ as a function of horizon size $R_e$ in hairy black hole 
solutions of $D=5$ ${\cal N}=8$ supergravity. The two curves correspond to 
solutions with two different AdS-invariant boundary conditions, 
namely $f=0$ (bottom) and $f=1$ (top).}
\label{3}
\end{figure}

\begin{figure}[htb]
\begin{picture}(0,0)
\put(42,254){$M_0$}
\put(411,19){$R_{e}$}
\end{picture}
\mbox{\epsfxsize=14cm \epsfysize=9cm \epsffile{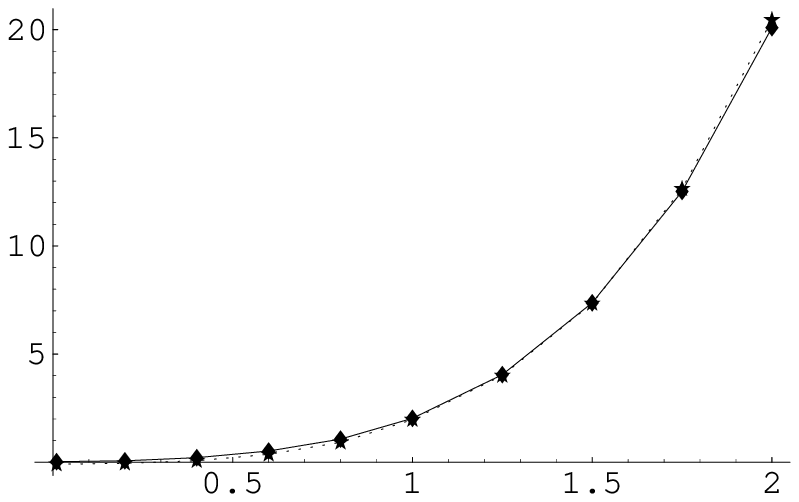}}
\caption{The integration constant $M_0$ as a function of horizon size $R_e$ 
in hairy black hole solutions of $D=5$ ${\cal N}=8$ supergravity.
The two curves correspond to solutions with two different AdS-invariant 
boundary conditions $f=0$ and $f=1$ (dotted line).}
\label{4}
\end{figure}

In Figure 1 we plot
the value $\phi_e$ of the field at the horizon of the hairy black holes 
as a function of horizon size
$R_e$. The two curves correspond to solutions with two 
different AdS-invariant boundary conditions, namely $f=0$ (bottom) and 
$f=1$ (top). Only for $f \rightarrow \infty$ we find no regular hairy 
black hole solutions. For all finite $f$ we find $\phi_e$ is nonzero
for all $R_e$, even for 
arbitrarily small black holes. This means the hairy black holes are 
disconnected from the Schwarschild-AdS solution. In Figure 2 we show the 
hair $\phi(r)$ of a black hole of size $R_e=.2$ that is a solution 
for boundary conditions corresponding to $f=1$. The hair $\phi(r)$ decays 
as $\ln (r)/r^2$ with a $1/r^2$ correction.
For given boundary conditions, the coefficient $\alpha$ in (\ref{hair5d})
fully characterizes the asymptotic profile of the hair.
Its value is shown in Figure 3 for a range of horizon sizes $R_e$, 
again for two different boundary conditions $f=0$ and $f=1$.
One sees that $\alpha$ reaches a minimum at
$R_e \approx .2$.

\begin{figure}[htb]
\begin{picture}(0,0)
\put(34,125){$E_h/3\pi^2$}
\put(204,12){$R_e$}
\put(250,148){$E_h/E_s$}
\put(455,12){$R_e$}
\end{picture}
\mbox{
\epsfxsize=7cm
\epsffile{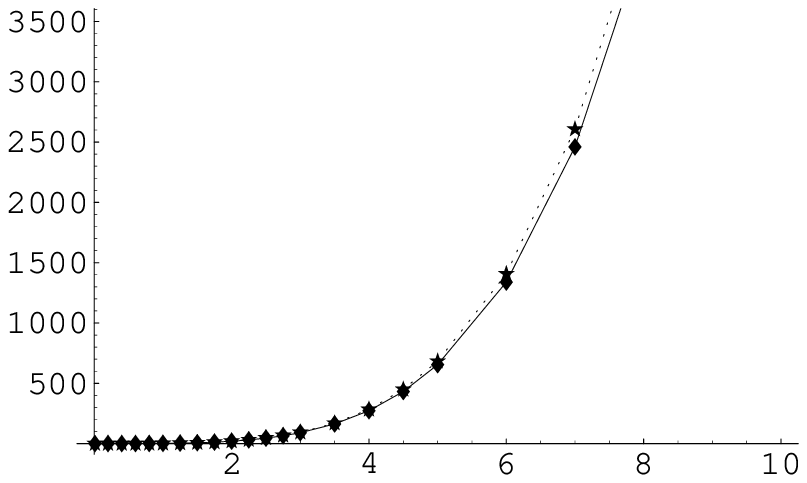}
\raisebox{2.3cm}{~~~~
\begin{minipage}{10cm}
\epsffile{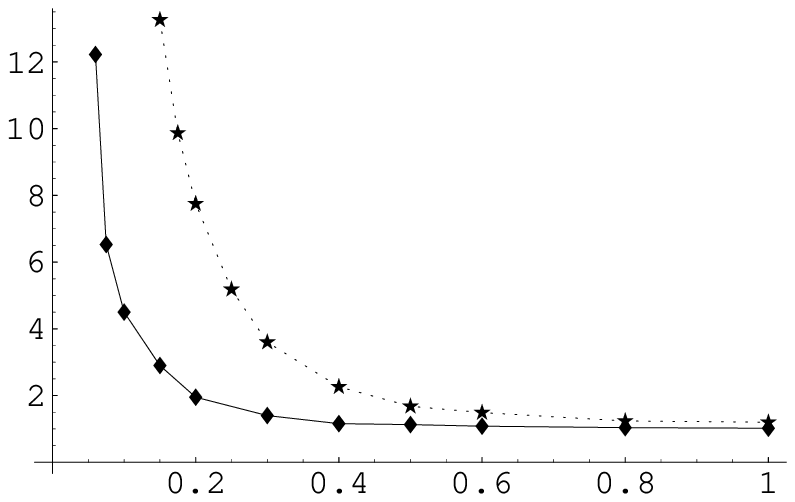}
\end{minipage}}}
\caption{left: The total mass $E_{h}/3\pi^2$ of hairy black holes as a 
function of horizon size $R_e$, in $D=5$ ${\cal N}=8$ supergravity with
two different AdS-invariant boundary conditions $f=1$ (top) and $f=0$ 
(bottom). 
right: The ratio $E_{h}/E_{s}$ as a function of horizon size $R_e$, where 
$E_s$ is the mass of a Schwarschild-AdS black hole of the same size $R_e$.}
\label{5}
\end{figure}

The integration constant $M_0$ as a function of horizon size $R_e$ is plotted
in Figure 4. Integrating the constraint equation (\ref{field5d2}) yields 
the following formal expression for $M_0$,
\beq
M_0 & = & \lim_{r \rightarrow \infty} \left[
e^{-{1\over 3} \int_{R_e}^{r}  d\hat r \ \hat r\pr^2}
(R_e^4 +R_e^2) \nonumber \right. \\
& & \left. +
\int_{R_e}^{r} e^{-{1\over 3}
\int_{\tilde r}^r  d\hat r \ \hat r\pr^2} 
\left[{2 \over 3}(V(\phi)-\Lambda) +{1 \over 3}
\left(1+ {\tilde r^2 \over \ell^2}\right)
\phi_{,\tilde r}^2 \right] \tilde r^{d-2} d\tilde r \right. \nonumber\\
& & \left. +\frac{2\alpha^2}{3}(\ln r)^2+\frac{2\alpha^2}{3}
\left(2f-\ln \alpha - \frac{\alpha}{2}\right)\ln r \right]
\eeq
One sees the hair exponentially 'screens' the Schwarschild-AdS mass. It also 
introduces new contributions to the gravitational mass which are absent in the
Schwarschild-AdS solutions. Figure 4 shows that $M_0 \sim R_e^4$ for 
large $R_e$.
For small $R_e$, however, we find $M_0<0$, at least 
with $f=1$ boundary conditions. As we explained above, 
this is not in conflict with the positive mass theorem \cite{Townsend84}
that is believed to hold because this only guarantees
the positivity of the conserved charge $Q[\partial_{t}]$ \cite{Hertog03b}.
The parameter $M_0$ therefore is of little physical significance. 
It is proportional to the finite gravitational contribution to 
$Q[\partial_{t}]$, but the total gravitational mass diverges. The relevant 
quantity is the total charge, which is given by
\be \label{mass5dhair}
E_{h} =Q[\partial_{t}]=2\pi^2 \left({3 \over 2}
M_0+\frac{1}{4}\alpha^2(\ln \alpha)^2
+\alpha^2\left(\frac{1}{4}-f\right)\ln \alpha
+\alpha^2\left(f^2-\frac{1}{2}f+\frac{1}{8}\right)\right).  
\ee

The mass $E_h$ is shown in Figure 5 as a function of horizon 
size $R_e$ and for two different boundary conditions $f=1$ (top) and 
$f=0$ (bottom). We find $E_h >0$ for all $R_e$ and for all boundary conditions
we have considered. For large $R_e$ one has $E_h \sim R_e^4$.
The mass is also compared with the mass $E_{s}$ of a Schwarschild-AdS black 
hole of the same size $R_e$. We find $E_{h}/E_{s} >1$ for all $R_e$ and 
$E_{h}/E_{s} \rightarrow 1$ for large $R_e$. Conversely it follows that
Schwarschild-AdS has always
larger entropy for a given mass. The ratio $E_{h}/E_{s}$ 
diverges for $R_e \rightarrow 0$. We find the hairy black 
holes can be arbitrarily small, but they only exist above a certain critical 
mass. The critical mass itself depends on the boundary conditions 
chosen. At the critical point the solution is nakedly singular.

For fixed AdS-invariant boundary conditions (with finite $f$), there is 
precisely one hairy black hole solution for a given total mass 
$Q[\partial_{t}]$ (larger than the critical mass). This is because regularity 
at the horizon (eq.(\ref{horcon})) uniquely determines the horizon size and 
the value of the scalar field at the horizon, for a given mass. This yields a 
unique combination of $\alpha$ and $M_0$, which are the two quantities that 
parameterize the class of static, sperically symmetric asymptotically AdS 
solutions. Thus we have found a one-parameter family of black holes with 
scalar hair, in a range of theories parameterized by $f$. Because 
Schwarschild-AdS is a solution too for all boundary conditions we have two 
very different black hole solutions for a given total mass, one with $\phi=0$ 
everywhere and one with nontrivial hair. So the scalar no hair theorem does not
hold in supergravity with asymptotically anti-de Sitter boundary conditions. 
Uniqueness is restored only for $f \rightarrow \infty$.

\subsection{Black Branes in Ten Dimensions}

$D=5\ \N=8$ supergravity is  believed to be a consistent truncation of
ten dimensional IIB supergravity on $S^5$. This means that it should be 
possible to lift our five dimensional solutions to ten dimensions.
Even though it is not known how to lift a general solution of $D=5,\ \N=8$
supergravity to ten dimensions, this is known for solutions that only involve 
the metric and scalars that saturate the BF bound \cite{Cvetic00}.
So we can immediately write down the ten dimensional analog of the hairy
black hole solutions. The ten dimensional solutions involve only the metric 
and the self dual five form. To describe them,
we first introduce coordinates on $S^5$ so that the metric on the unit sphere 
takes the form ($0 \le \xi \le \pi/2$)
\be
d\Omega_5 = d\xi^2 + \sin^2\xi d\varphi^2 + \cos^2\xi d\Omega_3
\ee
Letting $f = e^{\phi/2\sqrt3}$ and $\Delta^2 = f^{-2} \sin^2\xi + f \cos^2\xi$,
the full ten dimensional metric is 
\be
ds^2_{10} = \Delta ds_5^2 + f\Delta d\xi^2 + f^2\Delta^{-1} \sin^2\xi d\varphi^2
  + (f\Delta)^{-1} \cos^2\xi d\Omega_3
  \ee
This metric preserves an $SU(2)\times U(1)$ symmetry of the five sphere.
The five form is given by
\be
G_5 = - U \epsilon_5 - 3\sin\xi \cos\xi f^{-1} *df\wedge d\xi
\ee
where $\epsilon_5$ and $*$ are the volume form and dual in the five
dimensional solution and
\be
U= -2(f^2 \cos^2\xi +f^{-1}\sin^2\xi + f^{-1}).
\ee
One sees that the effect of the hair is to perturb the five sphere on the
horizon.

\setcounter{equation}{0}
\section{Hairy black holes in ${\cal N}=8$ $D=4$ Supergravity}

\subsection{AdS-invariant Boundary Conditions}

${\cal N}=8$ $D=4$ gauged supergravity \cite{deWit82} 
is the massless sector of the compactification of $D=11$ supergravity on 
$S^7$. We consider the truncation to its abelian $U(1)^4$ sector.
The resulting action is given by 
\be
\label{4-action}
S=\int d^4x\sqrt{-g}\left(\frac{R}{2}
-\frac{1}{2}\sum_{i=1}^3[(\nabla\phi_i)^2 
-2\cosh(\sqrt{2}\phi_i)] \right)+...
\ee
where the dots refer to fields that will be set to zero in our solutions.
We have also set $g^2=1/4$ so that the curvature scale of anti-de Sitter 
space is equal to one. The three remaining scalars decouple and they
each have mass
\be
m^2=-2,
\ee 
which lies in the range $ m^2_{BF} +1 >m^2 >m^2_{BF}$ in four dimensions.

We will consider solutions whose asymptotic behavior belongs to the following 
one-parameter class of AdS-invariant boundary conditions,
\be 
\label{4-scalar}
\phi_{i}(r,t,x^{a})=\frac{\alpha_{i}(t,x^a)}{r}+
\frac{f_{i} \alpha_{i}^2(t,x^a)}{r^2}
\ee
\beq 
\label{4-grr}
g_{rr}=\frac{1}{r^2}-\sum_{i=1}^{3}\frac{(1+\alpha^2_{i}/2)}{r^4}+
O(1/r^5) & \quad g_{tt}=-r^2 -1+O(1/r) \nonumber\\
g_{tr}=O(1/r^2) \qquad \qquad \qquad & \ \  \ \ \ g_{ab}= \bar g_{ab} +O(1/r) 
\nonumber\\
g_{ra} = O(1/r^2) \qquad \qquad \qquad & g_{ta}=O(1/r) \ \ 
\eeq
where $x^a=\theta,\phi$ and $f_i$ are constants without variation that label
the different boundary conditions. The conserved charges $Q=Q_\phi+Q_G$ that 
generate the asymptotic symmetries are finite and given by
\be
\label{charge4dN8}
Q[\xi]=Q_G[\xi]+{\lambda_{-} \over 2}\sum_{i=1}^3
\oint d\Omega_{d-2}\frac{\xi^\perp}{r}
r^{d-1}\left( \phi^2_{i} +{2f(\lambda_{+} - \lambda_{-}) \over d-1}
\phi^{\frac{d-1}{\lambda_-}}_{i} \right)
\ee

\subsection{Black Holes with Scalar Hair}

We now look for static, spherically symmetric black hole solutions with 
scalar hair that are asymptotically AdS.
We will concentrate on solutions in which only one scalar 
$\phi_1\equiv \phi$ is nonzero. Writing the metric as
\be
ds_4^2=-h(r)e^{-2\delta(r)}dt^2+h^{-1}(r)dr^2+r^2d\Omega_2^2
\ee
the field equations read
\be\label{hairy14d}
h\phi_{,rr}+\left(\frac{2h}{r}+\frac{r}{2}\phi_{,r}^2h+h_{,r}
\right)\phi_{,r}   =  V_{,\phi}
\ee
\be\label{hairy24d}
1-h-rh_{,r}-\frac{r^2}{2}\phi_{,r}^2h =  r^2V(\phi)
\ee

\begin{figure}[htb]
\begin{picture}(0,0)
\put(56,254){$\phi_e$}
\put(412,23){$R_{e}$}
\end{picture}
\mbox{\epsfxsize=14cm \epsfysize=9cm \epsffile{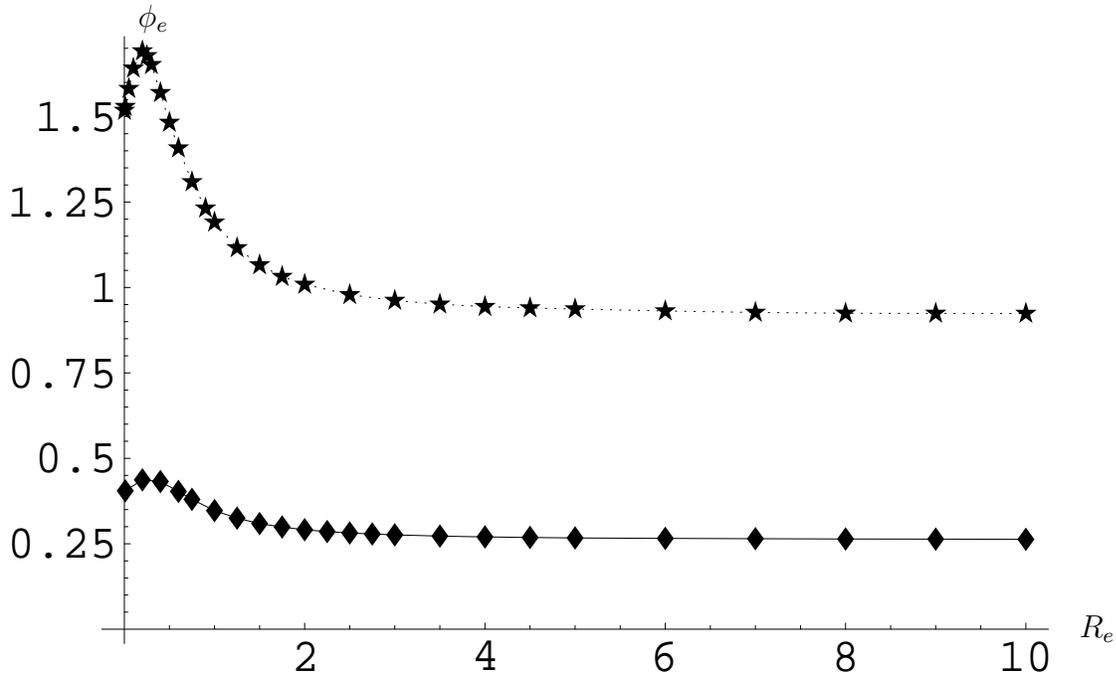}}
\caption{The scalar field $\phi_e$ at the horizon as a function of horizon size
$R_e$ in hairy black hole solutions of $D=4$ ${\cal N}=8$ supergravity.
The two curves correspond to solutions with two different AdS-invariant 
boundary conditions, namely $f=-1$ (bottom) and $f=-1/4$ (top).}
\label{6}
\end{figure}

Regularity at the event horizon $R_e$ imposes the constraint
\be \label{horcon4d}
\phi'(R_{e}) = {R_{e}V_{,\phi_{e}} \over 1-R_{e}^2V(\phi_{e})}
\ee
Asymptotic AdS invariance requires $\phi$ asymptotically decays as
\be \label{hair4d}
\phi(r)=\frac{\alpha}{r}+\frac{f\alpha^2}{r^2}, 
\ee
where $f$ is a given constant that is determined by the choice of
boundary conditions. Hence asymptotically
\be \label{asmetric4d}
h(r)=r^2+1+\alpha^2/2 -\frac{M_0}{r}, 
\ee
where $M_0$ is an integration constant.

\begin{figure}[htb]
\begin{picture}(0,0)
\put(34,125){$\phi$}
\put(204,15){$r$}
\put(250,141){$r\phi$}
\put(455,30){$r$}
\end{picture}
\mbox{
\epsfxsize=7cm
\epsffile{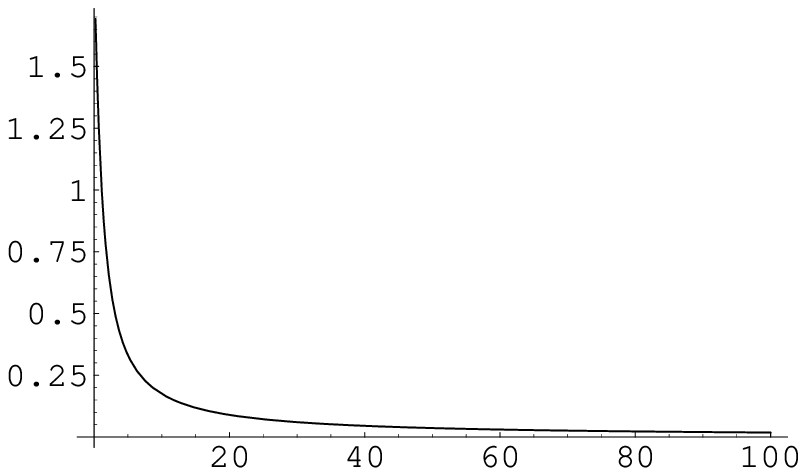}
\raisebox{2.3cm}{~~~~
\begin{minipage}{10cm}
\epsffile{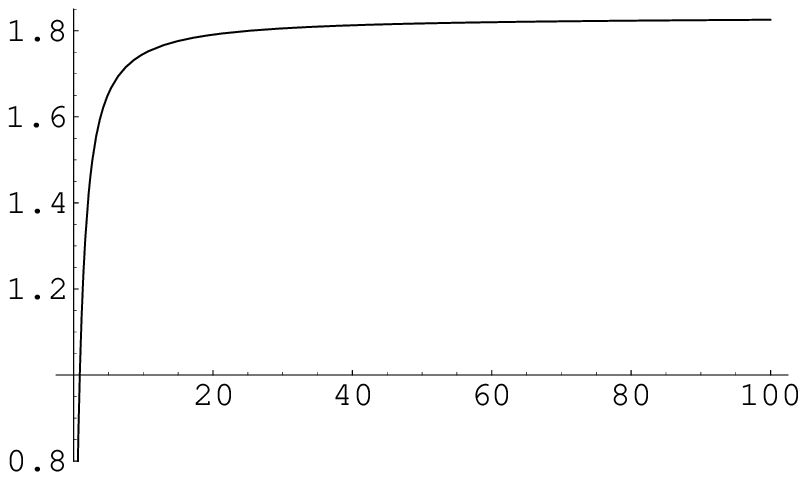}
\end{minipage}}}
\caption{The hair $\phi(r)$ (left) and $r\phi$ (right) outside a black hole 
of size $R_e=.2$, with boundary conditions specified by $f=-1/4$.}
\label{7}
\end{figure}

The Schwarschild-AdS black hole with $\phi=0$ everywhere 
outside the horizon is a solution for all AdS-invariant 
boundary conditions. The mass (\ref{charge4dN8}) reduces to
\be\label{mass4dschw}
Q[\partial_{t}]=4\pi M_0 =4\pi (R_e^3 +R_e),
\ee
which is the standard Schwarschild-AdS mass. 
However, numerical integration of the
field equations (\ref{hairy14d})-(\ref{hairy24d}) shows there is a large class
of boundary conditions that also admits a one-parameter family of static
spherically symmetric black hole solutions with scalar hair outside the 
horizon.

\begin{figure}[htb]
\begin{picture}(0,0)
\put(40,251){$\alpha$}
\put(412,20){$R_{e}$}
\end{picture}
\mbox{\epsfxsize=14cm \epsfysize=9cm \epsffile{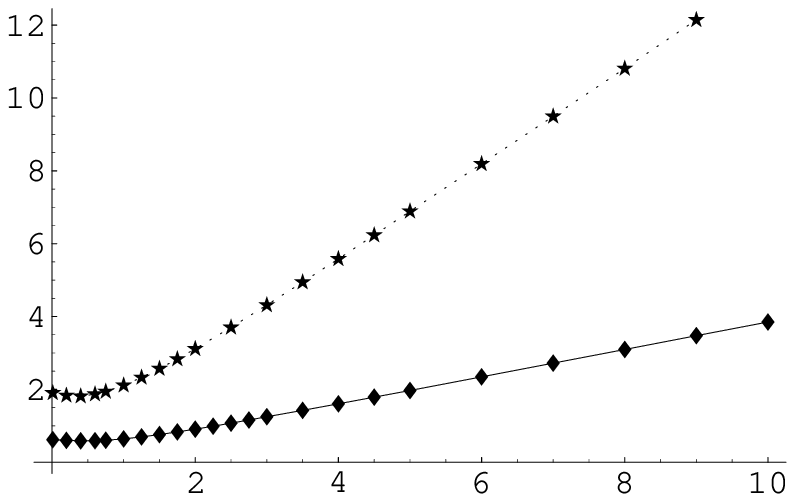}}
\caption{The coefficient $\alpha$ that characterizes the asymptotic profile 
of the hair $\phi(r)$ as a function of horizon size $R_e$ in hairy black hole 
solutions of $D=4$ ${\cal N}=8$ supergravity. The two curves correspond to 
solutions with two different AdS-invariant boundary conditions, 
namely $f=-1$ (bottom) and $f=-1/4$ (top).}
\label{8}
\end{figure}

\begin{figure}[htb]
\begin{picture}(0,0)
\put(42,252){$M_0$}
\put(412,18){$R_{e}$}
\end{picture}
\mbox{\epsfxsize=14cm \epsfysize=9cm \epsffile{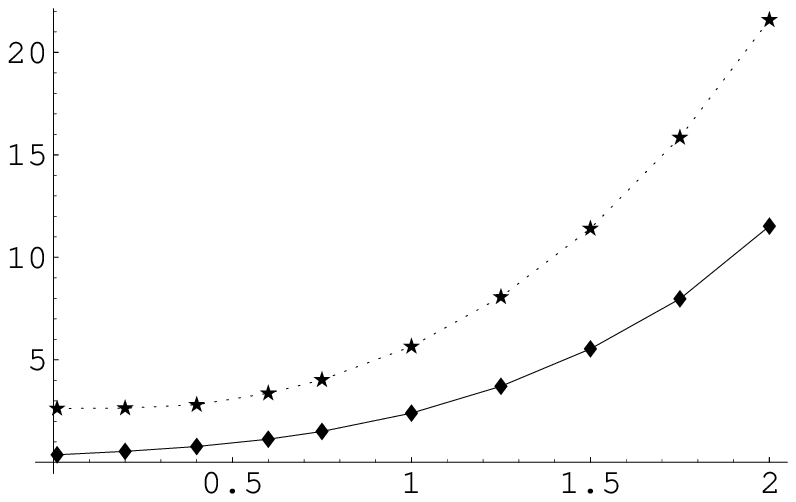}}
\caption{The integration constant $M_0$ as a function of horizon size $R_e$ 
in hairy black hole solutions of $D=4$ ${\cal N}=8$ supergravity.
The two curves correspond to solutions with two different AdS-invariant 
boundary conditions $f=-1$ (bottom) and $f=-1/4$ (top).}
\label{9}
\end{figure}

The value $\phi_e$ of the field at the horizon as a function of horizon size
$R_e$ is plotted in Figure 6. The two curves correspond to solutions with two 
different AdS-invariant boundary conditions, namely $f=-1$ (bottom) and 
$f=-1/4$ (top). Generically, we obtain $\phi_e>0$ if $f<0$ and $\phi_e<0$ 
for $f>0$. Only for $f=0$ and $f \rightarrow \infty$ we find no regular hairy 
black hole solutions. For all finite $f \neq 0$ we find $\phi_e$ is nonzero
for all $R_e$, even for 
arbitrarily small black holes. This means the hairy black holes are 
disconnected from the Schwarschild-AdS solution. In Figure 7 we show the 
hair $\phi(r)$ of a black hole of size $R_e=.2$ that is a solution 
for boundary conditions corresponding to $f=-1/4$. The hair $\phi(r)$ decays 
as $1/r$ with a $1/r^2$ correction.
For given boundary conditions, the coefficient $\alpha$ in (\ref{hair4d})
fully characterizes the asymptotic profile of the hair.
Its value is shown in Figure 8 for a range of horizon sizes $R_e$, 
again for two different boundary conditions $f=-1$ and $f=-1/4$.
One sees that $\alpha$ reaches a (different) positive minimum value at
$R_e \approx .2$. For large black holes, we have $\alpha \sim R_e$.

\begin{figure}[htb]
\begin{picture}(0,0)
\put(34,125){$E_h/4\pi$}
\put(204,14){$R_e$}
\put(245,148){$E_h/E_s$}
\put(455,12){$R_e$}
\end{picture}
\mbox{
\epsfxsize=7cm
\epsffile{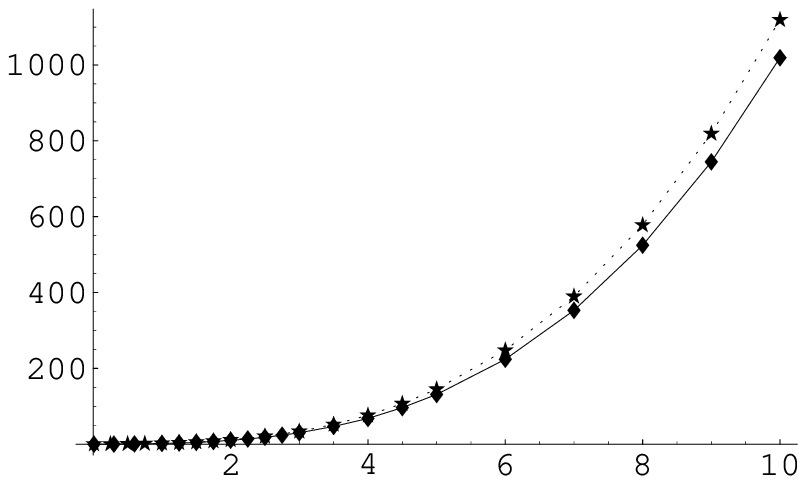}
\raisebox{2.3cm}{~~~~
\begin{minipage}{10cm}
\epsffile{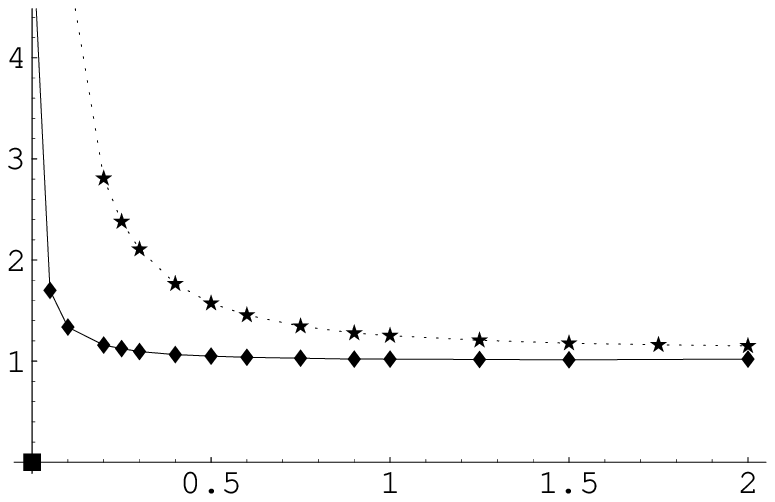}
\end{minipage}}}
\caption{ left: The total mass $E_{h}/4\pi$ of hairy black holes as a 
function of horizon size $R_e$, in $D=4$ ${\cal N}=8$ supergravity with
two different AdS-invariant boundary conditions $f=-1/4$ (top) and $f=-1$ 
(bottom). 
right: The ratio $E_{h}/E_{s}$ as a function of horizon size $R_e$, where 
$E_s$ is the mass of a Schwarschild-AdS black hole of the same size $R_e$}
\label{10}
\end{figure}
The integration constant $M_0$ as a function of horizon size $R_e$ is plotted 
in Figure 9. We find $M_0 \sim R_e^3$ for large $R_e$.
Integrating the constraint equation (\ref{hairy24d}) yields a formal
expression for $M_0$,
\beq
M_0 & = & \lim_{r \rightarrow \infty} \left[
e^{-{1\over 2} \int_{R_e}^{r}  d\hat r \ \hat r\pr^2}
(R_e^3 +R_e) \nonumber \right. \\
& & \left. +
\int_{R_e}^{r} e^{-{1\over 2}
\int_{\tilde r}^r  d\hat r \ \hat r\pr^2} 
\left[(V(\phi)+3) +{1 \over 2}
\left(1+ \tilde r^2 \right)
\phi_{,\tilde r}^2 \right] \tilde r^{2} d\tilde r 
+\frac{\alpha^2}{2}r \right]
\eeq
One sees the hair exponentially 'screens' the Schwarschild-AdS 
mass and introduces
new contributions to the gravitational mass which are absent in the 
Schwarschild-AdS solutions. 

The parameter $M_0$ is proportional to the finite gravitational contribution 
to the mass. It is, however, of little physical significance. Indeed the total
gravitational mass diverges. The relevant quantity is the total charge 
$Q[\partial_{t}]$, which is given by
\be \label{mass4dhair}
E_{h} =Q[\partial_{t}]=4\pi \left( M_0+\frac{4}{3}f\alpha^3 \right).  
\ee
The mass $E_h$ is shown in Figure 10 as a function of horizon 
size $R_e$ and for two different boundary conditions $f=-1/4$ (top) and 
$f=-1$ (bottom). We find $E_h >0$ for all $R_e$ and for all boundary conditions
we have considered. For large $R_e$ one has $E_h \sim R_e^3$.
The mass is also compared with the mass $E_{s}$ of a Schwarschild-AdS black 
hole of the same size $R_e$. We find $E_{h}/E_{s} >1$ for all $R_e$ and 
$E_{h}/E_{s} \rightarrow 1$ for large $R_e$. As before, $E_{h}$ is bounded 
from below - the hairy black hole solutions exist only above a certain 
critical mass. The critical mass itself depends on the boundary conditions 
chosen. At the critical point the solution is nakedly singular.

For fixed AdS-invariant boundary conditions, there is precisely one 
hairy black hole solution for a given total mass $Q[\partial_{t}]$. 
Hence the horizon size as well as the value of the scalar field at the horizon
are uniquely determined by $Q[\partial_{t}]$. Thus we have found a 
one-parameter family of black holes with scalar hair, in a range of theories
parameterized by $f$. Because Schwarschild-AdS is a solution too for all 
boundary conditions we have two very different black hole 
solutions for a given total mass, one with $\phi=0$ everywhere and 
one with nontrivial hair. So the scalar no hair theorem does not hold in 
$D=4$ ${\cal N}=8$ supergravity with asymptotically anti-de Sitter boundary 
conditions. Uniqueness is restored only in theories with $f=0$ or for 
$f \rightarrow \infty$. As before, the hairy black holes can be lifted to 
black branes in eleven dimensions with a perturbed $S^7$ on the horizon.

\setcounter{equation}{0}
\section{AdS/CFT with Generalized Boundary Conditions}

We have studied $D=5$ ${\cal N}=8$ supergravity, which is the low energy 
limit of string theory with $AdS_5 \times S^5$ boundary conditions, and
$D=4$ ${\cal N}=8$ supergravity, which is the low energy limit of string 
theory with $AdS_4 \times S^7$ boundary conditions. For these boundary 
conditions, the AdS/CFT correspondence \cite{Maldacena97} claims string 
theory is dual to a conformal field theory (CFT).
We have shown that the presence of scalars with sufficiently negative 
$m^2$ in both supergravity theories allows one to relax the boundary 
conditions on the metric and on the scalars to include non-localized matter 
distributions, while preserving the asymptotic $AdS$ symmetry group. 
According to the general AdS/CFT correspondence, there should be a dual 
conformal field theory corresponding to each choice of boundary conditions. 

Some aspects of AdS/CFT with generalized boundary conditions have already been
studied. Let us first consider $D=4$ ${\cal N}=8$ supergravity, for which 
the $AdS$-invariant boundary conditions were given in 
eqs.(\ref{4-scalar})-(\ref{4-grr}). For simplicity, we consider here 
generalizing the boundary conditions on a single scalar with $m^2 = -2$. 
Near the boundary, the field behaves as
\be \label{falloff}
\phi(t,r,x^a) = {\alpha (t,x^a) \over r} +{\beta (t,x^a) \over r^2}
\ee
where
\be \label{ba}
\beta = f \alpha^{2}
\ee
together with relaxed falloff conditions (\ref{4-grr}) on the metric.
\newline
Because
\be \label{range}
-{(d-1)^2\over 4} < m^2 < -{(d-1)^2\over 4}+1
\ee
it has been argued \cite{Balasubramanian98,Klebanov99} that $\phi$ can be 
associated with CFT operators of two possible dimensions,
\be
\Delta_{\pm} = { (d-1) \over 2} \pm \sqrt{{(d-1)^2\over 4}+m^2}
\ee
and that supersymmetry constraints are important to decide which assignment is
realized in each case. This goes back to the work of \cite{Breitenlohner82}, 
where it was argued there are two different AdS-invariant quantizations
of a scalar field with $m^2$ in the range (\ref{range}). For instance for
$m^2 = -2$ it is shown \cite{Breitenlohner82} one
can quantize the scalar field with a boundary condition $\alpha=0$, in which 
case it corresponds to an operator ${\cal O}'$ of dimension two in the 
boundary theory, or with a boundary condition $\beta =0$, in which case it 
corresponds to an operator ${\cal O}$ of dimension one in the boundary theory. 

However, this is a rather subtle issue. Indeed we have seen that  
with $\beta =0$ boundary conditions (which corresponds to choosing $f=0$ in 
(\ref{falloff4dphi})), 
one must weaken the falloff of the metric (\ref{falloff4dg}) in order 
to preserve asymptotic AdS invariance.
This means one does not really have two different quantum field theories 
on a given AdS background, but two quantum field theories on two
different AdS backgrounds. To quantize a test field
with $\beta =0$ on the standard AdS background is inconsistent because the 
gravitational backreaction would always diverge. By contrast, if one relaxes 
the asymptotic behavior 
of the metric, eq.(\ref{charge4d}) automatically yields a finite conserved
energy for both choices of boundary conditions. Neglecting backreaction for 
$f=0$ boundary conditions then amounts to neglecting the 
$O(1/r^5)$ correction to the $g_{rr}$-component in eq.(\ref{falloff4dg}).
Moreover, the two AdS-invariant quantizations we have 
discussed so far are only two members of a one-parameter family of 
AdS-invariant boundary conditions, parameterized by $f$. 
For all finite values of $f$ one must relax the falloff 
conditions on the metric components to ensure backreaction can be made small 
and the asymptotic symmetry group is preserved. 

This result explains the origin of the extra surface term in 
\cite{Breitenlohner82,Klebanov99}, which had to 
be added to the Euclidean action
to define a finite energy for the 'second' method of quantization.
It is concievable that the fact that the metric falls off slower than usual 
also explains other subtleties encountered in AdS/CFT which have to do with 
surface terms or normalization factors of correlation functions.

We now make some remarks on the relation between the dual field theories for
different $f$. For the `standard' boundary condition corresponding to 
$f \rightarrow \infty$ and $\alpha=0$, AdS/CFT relates 
$\phi$ in the boundary theory to a dimension two operator ${\cal O}'$ and
$\beta$ is interpreted as its expectation value. 
Boundary conditions with finite $f=1/f'^2$ are dual to certain 
deformations of the original CFT.
In particular, Witten has argued \cite{Witten02} they 
correspond to the addition of a term $W[{\cal O}']$, so that after 
formally replacing ${\cal O}'$ by its expectation value $\beta$ in $W$ one has
\be
\alpha = { \delta W \over \delta \beta}
\ee
For (\ref{ba}) this gives
\be \label{deform}
W= {2f'\over 3} \int d^3 x {\cal O}'^{3/2},
\ee
which indeed has the correct dimension to preserve conformal invariance.
For $f' \rightarrow \infty$, the boundary condition approaches the choice
$\beta=0$ which relates $\phi$ in the boundary theory to an operator 
${\cal O}$ of dimension one. Vice versa, the deformation that probes 
the family of boundary conditions starting with the CFT dual to $\beta=0$ 
boundary conditions is 
\be \label{deform2}
W = {f \over 3} \int d^3 x {\cal O}^3
\ee

It appears, therefore, that all AdS-invariant boundary conditions 
given in Section 2 can be incorporated in the AdS/CFT correspondence. 
The dual field theories differ from each other by multi-trace deformations 
that preserve conformal invariance. Thus we obtain a line of conformal 
fixed points. In theories with several scalar fields with $m^2$ in the range 
(\ref{range}) the different lines of conformal fixed points are parameterized
by the dimensionless constants $f_i$ that label the possible 
bulk boundary conditions. 

One expects, however, the change in the asymptotic behavior of the metric for 
finite $f$ should also deform the dual field theory. So 
presumably the full deformation should involve the CFT stress tensor as well. 
This point deserves further study and it may be particularly relevant to 
shed light on the deformations of 
${\cal N}=4$ super Yang-Mills that correspond to generalized boundary 
conditions in $D=5$ ${\cal N}=8$ supergravity. In this case, because the 
scalar fields saturate the BF bound, their asymptotic behavior (\ref{phi5d}) 
involves a logarithm for finite $f$. At first sight, it is not clear what 
could be the dual deformed CFT. From the bulk perspective however, the 
situation is rather similar to $D=4$ ${\cal N}=8$ supergravity - each scalar 
again gives rise to a one-parameter family of AdS-invariant boundary
conditions.

For scalars above the BF bound the total charge (\ref{charge4d}) reduces to 
the standard gravitational mass for localized matter fields 
(i.e. $f \rightarrow \infty$ boundary conditions). For those
boundary conditions it is known 
a positive mass theorem holds \cite{Gibbons83}. 
By contrast, it follows from (\ref{mass5dhair}) that for scalars saturating the
BF bound, there is a (finite) scalar contribution to the total charge
(\ref{charge5d}) even if $f \rightarrow \infty$. In this case the 
gravitational mass $M_0$ can be negative and need not be conserved during 
evolution \cite{Hertog03b}. Nevertheless, the positivity of the total charge
(\ref{charge5d}) is again ensured by the positive energy theorem 
\cite{Townsend84}.

The general proof \cite{Gibbons83,Townsend84} of the positive energy theorem 
in asymptotically AdS spaces relies on the existence of asymptotically 
supercovariant constant spinors. In a supergravity background such spinors - 
if they exist - will generate asymptotic global supersymmetry transformations.
Positivity of the energy is then an immediate consequence of the 
superalgebra \cite{Henneaux85}. It is an open question whether our generalized
boundary conditions are consistent with asymptotic supersymmetry.
From the dual field theory point of view, adding a multitrace 
interaction like (\ref{deform}) or (\ref{deform2}) breaks supersymmetry. 
However, because one must also take in account the effect of the
weakened metric boundary conditions on the CFT, this issue must be revisited.

\setcounter{equation}{0}
\section{Discussion}

${\cal N}=8$ supergravity theories in four and five dimensions contain scalar 
fields with masses in the range 
$-{(d-1)^2\over 4} \leq m^2 < -{(d-1)^2\over 4}+1$.
We have shown one can weaken the boundary conditions on the 
metric and on such scalars to include non-localized matter distributions
while preserving the asymptotic AdS symmetry group. The reason is that
the divergences of the gravitational charges are cancelled by contributions 
from the scalars, rendering the total charges finite. We find each scalar with
sufficiently negative $m^2$ gives rise to a one-parameter family of 
asymptotically AdS boundary conditions in which the metric falls off slower 
than usual. Generically the generators of the asymptotic symmetries also 
acquire additional finite contributions from the scalars.

For scalars above the BF bound, the finite scalar contributions 
vanish for `localized' matter distributions. Such configurations obey
boundary conditions $\phi \sim r^{-\lambda_{+}}$ (or faster) combined with
the standard falloff on all the metric components. In this case, the conserved
charge $Q[\partial_{t}]$ reduces to the standard gravitational mass, which is 
always finite and positive \cite{Gibbons83}. For all other AdS-invariant 
boundary conditions, 
including $\phi \sim r^{-\lambda_{-}}$, the scalars do contribute to the 
conserved charges and the boundary conditions on some metric components must 
be relaxed in order to preserve the asymptotic AdS symmetry group. 
The fact that the metric must fall off slower
than usual explains for instance the origin of the extra surface terms 
that are needed in AdS/CFT for these boundary conditions on scalar 
fields.

Scalars that saturate the BF bound yield finite (positive) contributions to 
the generators for {\em all} 
AdS-invariant boundary conditions. This even includes 
the usual `localized' matter distributions where $\phi \sim r^{-(d-1)/2}$ 
asymptotically, as was pointed out in \cite{Hertog03b}. The metric has the 
standard asymptotic behavior for those boundary conditions, but the 
gravitational mass - which appears in the metric - is generically neither 
positive nor conserved under evolution. 

For a single scalar, fixed AdS-invariant boundary conditions allow for a 
two-parameter family
of static, sperically symmetric asymptotic solutions. One parameter $\alpha$ 
characterizes the asymptotic profile of $\phi$ and a second parameter $M_0$ 
(together with $\alpha$) determines the asymptotic behavior of the metric. 
Relaxing the falloff on a single scalar in ${\cal N}=8$ supergravity  
in $D=4$ and $D=5$, we found there exists a one-parameter family of 
AdS black holes with scalar hair for all boundary conditions except a 
discrete number. 
The horizon size of the hairy black hole solutions as well as the value of 
the scalar field at the horizon are uniquely determined by a single charge,
namely the total mass 
$Q[\partial_{t}]$. The hairy black hole solutions only exist above a critical 
mass (which depends on the boundary conditions chosen), but their
horizon size can be arbitrarily small. Because the Schwarschild-AdS black hole
is also a solution for all boundary conditions, one has two very different 
black hole solutions for a given total mass (above a critical value).
Therefore the scalar no hair theorems do not hold in supergravity 
with asymptotically anti-de Sitter boundary conditions. 

For given boundary conditions, we find the scalar field at the horizon of 
the hairy black holes is always nonzero. Thus the hairy black holes 
are disconnected from the Schwarschild-AdS solution. We also find 
the hairy black holes are 
more massive than Schwarschild-AdS of the same size. The ratio of their 
masses, however, tends to one for large black holes. Conversely, for a given 
total mass, the Schwarschild-AdS black hole has always larger entropy. 

${\cal N}=8$ gauged supergravity in four dimensions is a consistent
truncation of M-Theory on $S^7$. Similarly ${\cal N}=8$ gauged supergravity 
in five dimensions is thought to be a consistent truncation of ten 
dimensional type IIB supergravity on $S^5$. Therefore our hairy black hole 
solutions can be lifted to new black brane solutions in ten or eleven 
dimensions. These black branes possess a horizon with a perturbed $S^5$ or 
$S^7$. We should mention, however, that the inhomogeneous black branes cannot 
be the endstate of the GL instability of Schwarschild-$AdS_5 \times S^5$, 
because this can only be seen in the dimensionally reduced 
setup if also the massive spin 2 fields corresponding 
to higher Kaluza-Klein modes of the metric are included.

The generalized boundary conditions can be incorporated in 
the AdS/CFT correspondence. Supergravity theories with different 
asymptotically AdS boundary conditions are dual to different CFT's. 
The dual field theories differ from each other by certain multi-trace 
interactions that preserve conformal invariance and by deformations involving 
the CFT stress tensor. Thus the range of possible
boundary conditions on each scalar 
defines a line of conformal fixed points on the gauge theory side. 
Whether all 
AdS-invariant boundary conditions given here are consistent with 
asymptotic supersymmetry remains an open question.

It would be interesting to study the hairy black holes in the context of the 
AdS/CFT correspondence. Large Schwarschild-AdS black holes have been 
conjectured to be described by an approximately thermal state in the gauge 
theory \cite{Witten98}. The existence of a second black hole 
solution with the same asymptotic charges poses a puzzle. It suggests there 
should be some observables in a `thermal' dual CFT state that are sensitive 
to the hair.

It would also be interesting to study cosmic censorship \cite{Penrose69}
in anti-de Sitter space \cite{Hertog03b,Hertog03a,Hertog03} 
with generalized boundary conditions. In \cite{Hertog03b} initial data 
were constructed in ${\cal N}=8$ $D=5$ supergravity in which the 
scalar in (\ref{lagr}) decays as $\ln r/r^2$ outside a central homogeneous 
region where $\phi =\phi_0$. A large radius cutoff was imposed to render the 
gravitational mass finite. The mass of the initial data was then compared 
with the mass needed to form a black hole large enough to enclose the 
singularity that develops in the central region. Of course, the cutoff 
destroys 
the asymptotic symmetries. However, using our results one can now remove the 
cutoff and repeat the analysis of \cite{Hertog03b} while preserving asymptotic
AdS-invariance. In addition, one can generalize the analysis to theories and
initial data involving 
scalar fields with different $m^2$ in the range (\ref{range}). Generically the
`gravitational' mass $M_0$ will not be conserved during evolution. Instead the
relevant quantity to decide whether or not low mass initial data of this type 
can evolve to black holes is the total conserved charge $Q[\partial_{t}]$.

\bigskip

\centerline{{\bf Acknowledgments}}
We thank M. Henneaux, C. Herzog, G. Horowitz, T. Okamura, H. Reall 
and R. Roiban for useful discussions. 
This work was supported in part by NSF grant PHY-0244764.

\end{document}